\documentclass[prd,twocolumn,twoside,preprintnumbers,superscriptaddress,nofootinbib]{revtex4}
\usepackage{amsmath,slashed}
\usepackage{graphicx,graphics,color}
\usepackage{dcolumn}
\usepackage[hyperfootnotes=false]{hyperref}
\usepackage{xspace}

\newcommand{\Order}{\mathcal{O}}
\newcommand{\MeV}{\,\text{MeV}}
\newcommand{\GeV}{\,\text{GeV}}
\newcommand{\TeV}{\,\text{TeV}}

\newcommand{\Lagr}{\mathcal{L}}
\newcommand{\beq}{\begin{equation}}
\newcommand{\eeq}{\end{equation}}
\renewcommand{\Re}{\text{Re}\,}
\renewcommand{\Im}{\text{Im}\,}
\newcommand{\Br}{\text{Br}}
\newcommand{\ecm}{e\,\text{cm}}

\allowdisplaybreaks[1]

\begin{document}

\preprint{INT-PUB-18-039, PSI-PR-18-09}
\title{Combined explanations of $\boldsymbol{(g-2)_{\mu,e}}$ and implications for a large muon EDM}

\author{Andreas Crivellin}
\affiliation{Paul Scherrer Institut, CH--5232 Villigen PSI, Switzerland}
\author{Martin Hoferichter}
\affiliation{Institute for Nuclear Theory, University of Washington, Seattle, WA 98195-1550, USA}
\author{Philipp Schmidt-Wellenburg}
\affiliation{Paul Scherrer Institut, CH--5232 Villigen PSI, Switzerland}

\begin{abstract}
With the long-standing tension between experiment and Standard-Model (SM) prediction in the anomalous magnetic moment of the muon, $a_\mu=(g-2)_\mu/2$, at the level of $3$--$4\sigma$, it is natural to ask if there could be a sizable effect in the electric dipole moment (EDM) $d_\mu$ as well. In this context it has often been argued that in UV complete models the electron EDM, which is very precisely measured, excludes a large  effect in $d_\mu$. However, the recently observed $2.5\sigma$ tension in $a_e=(g-2)_e/2$, if confirmed, requires that the muon and electron sectors effectively decouple to avoid constraints from $\mu\to e\gamma$. We briefly discuss UV complete models that possess such a decoupling, which can be enforced by an Abelian flavor symmetry $L_\mu-L_\tau$. We show that, in such scenarios, there is no reason to expect a correlation between the electron and muon EDM, so that the latter can be sizable. New limits on $d_\mu$ improved by up to two orders of magnitude are expected from the upcoming $(g-2)_\mu$ experiments at Fermilab and J-PARC. Beyond, a proposed dedicated muon EDM experiment at PSI could further advance the limit. In this way, future improved measurements of $a_e$, $a_\mu$, as well as the fine-structure constant $\alpha$ are not only set to provide exciting precision tests of the SM, but, in combination with EDMs, to reveal crucial insights into the flavor structure of physics beyond the SM.
\end{abstract}

\maketitle

\section{Introduction}

Ever since Schwinger's seminal prediction $a_\ell=\alpha/(2\pi)$~\cite{Schwinger:1948iu}, magnetic moments of charged leptons have served as powerful precision tests first of quantum electrodynamics (QED) and later of the full SM. In fact, for the muon there exists a tantalizing tension between the measurement~\cite{Bennett:2006fi}
\beq
a_\mu^\text{exp}=116,\!592,\!089(63)\times 10^{-11}
\eeq
(corrected for the updated ratio of proton and muon magnetic moments~\cite{Mohr:2015ccw}) and the SM prediction. The latter is currently being re-evaluated in a community-wide effort prompted by upcoming improved measurements at Fermilab~\cite{Grange:2015fou} and J-PARC~\cite{Saito:2012zz} (see also~\cite{Gorringe:2015cma}), with promising recent advances in hadronic vacuum polarization~\cite{Chakraborty:2016mwy,Jegerlehner:2017lbd,DellaMorte:2017dyu,Davier:2017zfy,Borsanyi:2017zdw,Blum:2018mom,Keshavarzi:2018mgv,Colangelo:2018mtw}, hadronic light-by-light scattering~\cite{Colangelo:2015ama,Green:2015sra,Gerardin:2016cqj,Blum:2016lnc,Colangelo:2017qdm,Colangelo:2017fiz,Blum:2017cer,Hoferichter:2018dmo,Hoferichter:2018kwz}, and higher-order hadronic corrections~\cite{Kurz:2014wya,Colangelo:2014qya}. Current evaluations point towards a discrepancy 
\beq
\label{Delta_amu}
\Delta a_\mu=a_\mu^\text{exp} - a_\mu^\text{SM}\sim 270(85)\times 10^{-11}
\eeq
of around $3$--$4\sigma$ (for definiteness, we choose a value at the lower end). 

This tension raises the question about the existence of effects beyond the SM (BSM) in the EDM of the muon. Here, the present EDM bound is~\cite{Bennett:2008dy}
\beq
\label{dmulimit}
|d_\mu|<1.5\times 10^{-19}\ecm\qquad 90\% \,\text{C.L.},
\eeq
which is about 600 times larger than expected from the central value of $a_\mu$ assuming that the imaginary part of the corresponding BSM contribution is as large as the real one. 
In contrast, the electron EDM is very precisely measured~\cite{Baron:2013eja,Andreev:2018ayy} with an upper limit of
\beq
\label{delimit}
|d_e|<1.1\times 10^{-29}\ecm\qquad 90\% \,\text{C.L.},
\eeq
which indicates a very small or even vanishing phase of any BSM contribution. Models with minimally-flavor-violating (MFV) structures~\cite{Chivukula:1987fw,Hall:1990ac,Buras:2000dm,DAmbrosio:2002vsn,He:2014uya} then predict $d_\mu=m_\mu/m_e d_e$, leading to
\beq
\label{dmu_MFV}
|d_\mu^\text{MFV}|<2.3\times 10^{-27}\ecm\qquad 90\% \,\text{C.L.}
\eeq
This is eight orders of magnitude below the current limit, but it is imperative to keep in mind that it is derived under the strong assumption of MFV. 

MFV is strongly challenged by recent experimental measurements in semileptonic $B$ meson decays (see~\cite{Crivellin:2018gzw} for a recent review)  and by a new, indirect, measurement of $a_e$. Until recently, the direct measurement of $a_e$~\cite{Hanneke:2008tm}
\beq
a_e^\text{exp}=1,\!159,\!652,\!180.73(28)\times 10^{-12}
\eeq
agreed with the SM prediction~\cite{Aoyama:2017uqe} 
\beq
a_e^\text{SM}\big|_{\alpha_\text{Rb}}=1,\!159,\!652,\!182.03(72)\times 10^{-12},
\eeq
derived from the fine-structure constant as measured in Rb atomic interferometry~\cite{Bouchendira:2010es}, at the level of $1.7\sigma$, 
with the uncertainty completely dominated by $\Delta a_e^\text{SM}$, i.e.\ limited by the precision of the Rb measurement of $\alpha$. This situation changed significantly with a new measurement of $\alpha$ using Cs atoms~\cite{Parker:2018}, implying
\beq
a_e^\text{SM}\big|_{\alpha_\text{Cs}}=1,\!159,\!652,\!181.61(23)\times 10^{-12}.
\eeq
Thus
\beq
\label{Delta_ae}
\Delta a_e=a_e^\text{exp} - a_e^\text{SM}=-0.88(36)\times 10^{-12},
\eeq
which corresponds to a $2.5\sigma$ deviation, at a level of accuracy improved by a factor of $2$.\footnote{The extraction of $\alpha$ from atomic interferometry relies on the Rydberg constant $R_\infty$ from ~\cite{Mohr:2015ccw}, with a quoted uncertainty of $6$ ppt. Since a shift in $R_\infty$ could be a possible resolution of the proton radius puzzle~\cite{Pohl:2010zza}, 
one might wonder about the impact on the determination of $\alpha$, but the suggested shift $\Delta R_\infty/R_\infty=-0.03$ ppb translates to 
$\Delta a_e^\text{SM}=\Delta R_\infty/R_\infty\, a_e^\text{SM}/2=-0.018\times 10^{-12}$, in the right direction, but a factor $50$ too small to explain~\eqref{Delta_ae}.} 
Most crucially, the sign of $\Delta a_e$ is opposite to $\Delta a_\mu$, contradicting the MFV hypothesis.
It also excludes a resolution of $\Delta a_\mu$ in terms of a dark photon, leading to a positive sign, at $99\%$ confidence level~\cite{Parker:2018}, while a new axially-coupled light degree of freedom would result in the negative sign required by $\Delta a_e$~\cite{Kahn:2016vjr}, but could not accommodate $\Delta a_\mu$ at the same time. 

Given the deviation~\eqref{Delta_amu}, primary attention has focused on $a_\mu$, but interest in $a_e$ as a test of QED and the full SM goes back decades, see e.g.~\cite{Brodsky:1980zm};
more recently, the role of $a_e$ as a precision test of the SM has been studied in~\cite{Girrbach:2009uy,Giudice:2012ms}.
Starting from the benchmark that BSM contributions scale with the square of the lepton mass, one would expect $\Delta a_e\sim 0.06(2)\times 10^{-12}$, thus another factor of $5$ below current sensitivities. To reach that level of precision, concurrent improvements both in the direct measurement $a_e^\text{exp}$ and $\alpha$ are clearly necessary, but also the sub-leading uncertainties from the numerical integration error in the $4$- and $5$-loop QED coefficients were found to be relevant~\cite{Giudice:2012ms}. Since the semi-analytical work by Laporta~\cite{Laporta:2017okg} eliminates the $4$-loop uncertainty completely, while the improved $5$-loop results from~\cite{Aoyama:2017uqe} push the remaining uncertainty to the same level as hadronic corrections $\lesssim 0.02\times 10^{-12}$, such improved measurements can now be translated directly into yet more stringent SM precision tests.

From these considerations the emergence of a non-zero $\Delta a_e$, in particular the opposite sign, would be surprising, and a BSM explanation almost necessarily has to violate the quadratic mass scaling. 
Such a scenario itself is not entirely unexpected~\cite{Giudice:2012ms}, given that only one power of the mass is coming from the equations of motions, while the second one results from assuming a SM-like structure of the Yukawa interactions. Therefore, some enhancement with respect to the MFV mass scaling is necessary to explain $a_e$. One possible model displaying such an enhancement, based on a light scalar, has been recently proposed in~\cite{Davoudiasl:2018fbb}.

In this paper, we stress that a common feature that emerges in explanations along these lines concerns an effective decoupling of the $\mu$ and $e$ BSM sectors, due to the stringent limits from $\mu\to e\gamma$. We discuss explicit models which possess such a decoupling, and, by comparing $g-2$ to EDM limits, we argue that this decoupling allows for a large muon EDM despite the 
stringent electron EDM limit. Such scenarios could be probed at the upcoming $(g-2)_\mu$ experiments at Fermilab and J-PARC, and, potentially, a dedicated muon EDM experiment at PSI.

\section{EFT analysis}

We start by collecting the relevant expressions for magnetic moments and $\mu\to e\gamma$. The effective Hamiltonian
\beq
\label{HeffLFV}
{\cal{H}}_\text{eff}= c^{\ell_{f}\ell_{i}}_{R} \, \bar{\ell}_{f}
\sigma_{\mu \nu} P_{R} \ell_{i} F^{\mu \nu}+\text{h.c.}
\eeq
gives
\begin{align}
\label{Brmuegamma}
a_{\ell_i}&= -\frac{2m_{\ell_{i}}}{e}\, \big(c^{\ell_{i}\ell_{i}}_{R}+c^{\ell_{i}\ell_{i}*}_{R}\big)=-\frac{4m_{\ell_{i}}}{e}\, \Re c^{\ell_{i}\ell_{i}}_{R}, \notag\\
d_{\ell_i} &= i\big(c^{\ell_{i}\ell_{i}}_{R}-c^{\ell_{i}\ell_{i}*}_{R}\big)=-2\,\Im c^{\ell_{i}\ell_{i}}_{R},\notag\\ 
\Br[\mu \to e \gamma]&=\frac{m_{\mu}^3}{4\pi \, \Gamma_{\mu}} \big(|c^{e\mu}_{R} |^{2}+ |c^{\mu e}_{R} |^{2}\big),
\end{align}
where $\ell_i,\ell_f\in\{e,\mu,\tau\}$. This decomposition emphasizes that in general there are no correlations between magnetic moments and lepton flavor violation, such correlations are always model dependent~\cite{Lindner:2016bgg}. Furthermore, in the definition of the Wilson coefficients~\eqref{HeffLFV} we did not implicitly assume MFV, i.e.\ $a_\mu$ and $a_e$ are linear (rather than quadratic) in $m_\mu$ and $m_e$, respectively.

Therefore, for generic BSM scenarios the effect in muons is larger than in electrons
\beq
\label{size}
-3.0\gtrsim\Re c^{\mu\mu}_{R}/\Re c^{ee}_{R}\gtrsim-130,
\eeq
varying both $a_\mu$ and $a_e$ within their preferred $2\sigma$ ranges, with a central value $\Re c^{\mu\mu}_{R}/\Re c^{ee}_{R}\sim-15$.
Likewise, the central values in~\eqref{Delta_amu} and~\eqref{Delta_ae} give
\beq
\label{limits_phase}
\bigg|\frac{\Im c^{ee}_{R}}{\Re c^{ee}_{R}}\bigg|\lesssim 6\times 10^{-7},\qquad
\bigg|\frac{\Im c^{\mu\mu}_{R}}{\Re c^{\mu\mu}_{R}}\bigg|\lesssim 600.
\eeq
Thus, the phase of $c_R^{ee}$ must be very small, while that of $c_R^{\mu\mu}$ is largely unconstrained. The future $(g-2)_\mu$ experiments will be sensitive to $|d_\mu|\sim 10^{-21}\ecm$~\cite{Gorringe:2015cma}, but probing phases around $45^{\circ}$ requires yet another order of magnitude improvement.

\begin{figure}
\centering
\includegraphics[width=0.5\linewidth]{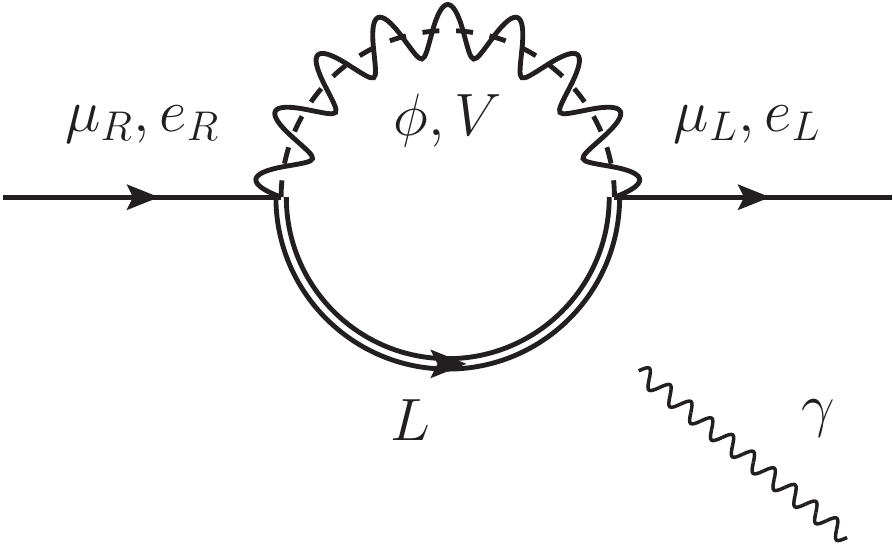}
\caption{Generic $1$-loop diagram contributing to the dipole operator with fermion $L$ and scalar or vector $\phi$ or $V$, respectively.}
\label{fig:muebound}
\end{figure}

Finally, the EFT analysis shows that solutions where the BSM particles couple to muons and electrons simultaneously
$c^{e\mu}_{R}=\sqrt{c^{ee}_{R}c^{\mu\mu}_{R}}$ are excluded since the resulting
\beq
\Br[\mu \to e \gamma]  = \frac{\alpha m_\mu ^2}{16 m_e\Gamma _\mu }|\Delta a_\mu \Delta a_e|\sim 8\times 10^{-5}
\label{muegamma}
\eeq
violates the MEG bound~\cite{TheMEG:2016wtm}
\beq
\Br[\mu \to e \gamma]<4.2\times 10^{-13}\qquad 90\% \,\text{C.L.}
\eeq
by $8$ orders of magnitude. The relation~\eqref{muegamma} arises in minimal models where the muon and electron sector are not decoupled, as shown in Fig.~\ref{fig:muebound}, e.g.\ it holds if a single new heavy fermion $L$ is added to the SM. Furthermore, the bound still applies if a new scalar or vector is introduced in addition. Therefore, to evade the bound one needs at least two (new) fermions or two new scalars/vectors in the loop. This includes the case of a single new scalar/vector coupling flavor-diagonally to muons and electrons. However, in such a scenario no chiral enhancement is possible.  
 
In this way, the spectacular failure~\eqref{muegamma} already indicates that much more intricate constructions are necessary to obtain a viable model that displays chiral enhancement, i.e.\ the muon and the electron sector must be separated. We now turn to models which can possess such a decoupling of the two sectors.

\section{Models explaining both anomalies}

In order to explain the quite large effect in $a_\mu$, which is of the order of the electroweak (EW) contribution in the SM, and the relatively even larger effect in $a_e$, any viable BSM mechanism requires some form of enhancement: it either has to be light, has to have $\Order(1)$ couplings for TeV-scale masses, or it needs to possess a chiral enhancement, i.e.\ a coupling to the Higgs field much larger than the SM one $m_\ell/v$. An example for such a chiral enhancement is $\tan\beta$ in the MSSM or $m_q/m_\ell$ in models with leptoquarks (LQs). The necessity of an enhanced Higgs coupling in any model realized above the EW breaking scale can be easily understood by looking at the gauge-invariant effective operators 
$Q_{eW}^{fi}=\bar \ell_f \sigma^{\mu\nu} \tau^I\ell_i H W^I_{\mu\nu} $ and $Q_{eB}^{fi}=\bar \ell_f \sigma^{\mu\nu} \ell_i H F_{\mu\nu} $, which explicitly involve the SM Higgs doublet~\cite{Buchmuller:1985jz,Grzadkowski:2010es}. 

As mentioned in the introduction, light (pseudo-) vector particles (``dark photons'') are problematic. Neutral vectors give a necessarily positive effect and can therefore only account for $a_\mu$, while neutral axial vectors give a negative effect and are therefore only compatible with $a_e$. Furthermore, the preferred regions from $a_\mu$ and $a_e$ are in general in tension with other constraints~\cite{Kahn:2016vjr}. While a light scalar, as proposed in~\cite{Davoudiasl:2018fbb}, provides in principle a relatively economical solution, the model is not yet UV complete and its UV completion again requires heavy BSM degrees of freedom.  
Here, we will instead consider models realized above the EW breaking scale with chiral enhancement. In general, respecting Lorentz invariance and renormalizability, one can only add new scalar, vector, 
and/or fermion fields to the particle content of the SM.

\subsection{New scalars, vectors, and fermions}

Without any assumptions on the specific model, the coupling of fermions to SM leptons and scalars/vectors can be parametrized as
\begin{align}
\label{Lagr_general}
\Lagr_\Phi&= \bar \Psi \left( {\Gamma _{\Psi \Phi }^{iL}{P_L} + \Gamma _{\Psi \Phi }^{iR}{P_R}} \right){\ell _i}{\Phi ^*} +\text{h.c.},\\
\Lagr_{V}&= \bar \Psi \left( {\Gamma _{\Psi V }^{iL}\gamma^\mu{P_L} + \Gamma _{\Psi V }^{iR}\gamma^\mu{P_R}} \right){\ell _i}{V_\mu ^*} +\text{h.c.},\label{Lagr_vector}
\end{align}
where a sum over all fermions $\Psi$ and scalars (vectors) $\Phi$ ($V^\mu$) is implicitly understood. With these conventions at hand the contribution of new scalars and vectors to dipole moments can be written as (see also~\cite{Freitas:2014pua,Stockinger:2006zn})
\begin{align}
c_{R\Phi}^{fi} &= \frac{e}{{16{\pi ^2}}}\Gamma _{\Psi \Phi }^{fL*}\Gamma _{\Psi \Phi }^{iR}{M_\Psi }\frac{f_\Phi\big(\frac{M_\Psi^2}{M_\Phi^2}\big)+Q g_\Phi\big(\frac{M_\Psi^2}{M_\Phi^2}\big)}{M_\Phi^2}\notag\\
&+ \frac{e}{{16{\pi ^2}}}\big(m_{\ell_i}\Gamma _{\Psi\Phi }^{fL*}\Gamma _{\Psi\Phi }^{iL}+m_{\ell_f}\Gamma _{\Psi\Phi }^{fR*}\Gamma _{\Psi\Phi }^{iR}\big)\notag\\
&\qquad\times\frac{\tilde f_\Phi\big(\frac{M_\Psi^2}{M_\Phi^2}\big)+Q \tilde g_\Phi\big(\frac{M_\Psi^2}{M_\Phi^2}\big)}{M_\Phi^2} ,\label{c_R_general_Phi}\\
c_{R V}^{fi} &= \frac{e}{{16{\pi ^2}}}\Gamma _{\Psi V }^{fL*}\Gamma _{\Psi V }^{iR}M_\Psi \frac{{{f_V}\big(\frac{M_\Psi^2}{M_V^2}\big) + Q{g_V}\big(\frac{M_\Psi^2}{M_V^2}\big)}}{{M_V ^2}}\notag\\
&+\frac{e}{{16{\pi ^2}}}\big(m_{\ell_i}\Gamma _{\Psi V}^{fL*}\Gamma _{\Psi V}^{iL}+m_{\ell_f}\Gamma _{\Psi V}^{fR*}\Gamma _{\Psi V}^{iR}\big)\notag\\
&\qquad\times\frac{{{\tilde f_V}\big(\frac{M_\Psi^2}{M_V^2}\big) + Q{\tilde g_V}\big(\frac{M_\Psi^2}{M_V^2}\big)}}{{M_V ^2}},\label{c_R_general_V}
\end{align}
with
\begin{align}
f_\Phi(x)&=2\tilde g_\Phi(x)=\frac{x^2-1-2x\log x}{4(x-1)^3},\notag\\
g_\Phi(x)&=\frac{x-1-\log x}{2(x-1)^2},\notag\\
\tilde f_\Phi(x)&=\frac{2x^3+3x^2-6x+1-6x^2\log x}{24(x-1)^4},\notag\\
f_V(x) &= \frac{x^3-12x^2 + 15x -4 + 6x^2\log x}{4(x - 1)^3},\notag\\
g_V(x) &= \frac{x^2 - 5x +4 + 3x\log x}{2(x - 1)^2},\notag\\
\tilde f_V(x) &= \frac{-4x^4+49x^3-78x^2 + 43x -10 - 18x^3\log x}{24(x - 1)^4},\notag\\
\tilde g_V(x) &= \frac{-3(x^3-6x^2 + 7x -2 + 2x^2\log x)}{8(x - 1)^3},\label{loop_functions}
\end{align}
where $Q$ is the electric charge of the fermion.
We calculated the contribution of the massive vector boson in unitary gauge, so that the effects of Goldstone bosons are automatically included, which is possible since the matching on dipole operators gives a finite result. The terms proportional to the heavy fermion mass are the ones that can be chirally enhanced.
These contributions have an arbitrary phase also for $i = f$ while, due to Hermiticity of the Lagrangian, the terms which are not chirally enhanced, i.e.\ proportional to $\Gamma _{\Psi V,\Phi }^{fL*}\Gamma _{\Psi V,\Phi }^{iL}$ and $\Gamma _{\Psi V,\Phi }^{fR*}\Gamma _{\Psi V,\Phi }^{iR}$ (included here for completeness), are real for flavor-conserving dipole transitions. Note that the relations~\eqref{Lagr_general}--\eqref{c_R_general_V} are not manifestly $SU(2)$ invariant but only invariant with respect to $U(1)_\text{EM}$. Therefore, we will illustrate them for several (simplified) models: LQs, the minimal supersymmetric SM (MSSM), Little-Higgs-inspired models, and, in more detail, a simplified model with new heavy leptons (and possibly a new scalar).

\subsection{Specific models}

{\bf Leptoquarks}
\newline
In LQ models one adds in their minimal version only one new scalar or vector particle to the SM. Therefore, they are, with respect to their particle content, minimal models with chiral enhancement. In constructing these models one demands that the couplings to quarks and leptons respect SM gauge invariance, resulting in $5$ vector LQs and $5$ scalars LQs~\cite{Buchmuller:1986zs}. Therefore, in~\eqref{c_R_general_Phi} and \eqref{c_R_general_V} $M_\Psi$ corresponds to the quark mass, $M_\Phi$ and $M_V$ to the LQ mass, respectively, and a factor $N_c=3$ has to be added to take into account the fact that quarks and LQs are colored. 
There are two representations of scalar LQs that can easily accommodate $a_\mu$ via a chiral enhancement by the top mass~\cite{Djouadi:1989md,Davidson:1993qk,Couture:1995he,Chakraverty:2001yg,Bauer:2015knc,Das:2016vkr,Biggio:2016wyy,ColuccioLeskow:2016dox} and two vector LQs whose effect in dipole moments can be enhanced by the bottom mass~\cite{ColuccioLeskow:2016dox}. Therefore, even for TeV-scale masses, one can easily explain the tension in $a_\mu$ or $a_e$ for couplings of order $0.1$.

However, in their minimal version, LQs are a single-particle extension of the SM and thus subject to the constraint in~\eqref{muegamma}. Therefore, they can only account for $a_\mu$ by decoupling the electron sector completely,\footnote{Decoupling of the electron sector is also motivated by the anomalies in $R(K)$ and $R(K^*)$, where $\mu\to e\gamma$ requires small or vanishing couplings to electrons~\cite{Crivellin:2017dsk}.} and thus cannot explain $a_e$ at the same time. However, this implies that also the EDMs are decoupled and the phase of $c_R^{\mu\mu}$ is not subject to any serious constraint, so that one would expect naturally $|d_\mu|\sim e/(2m_\mu) \Delta a_\mu \sim 3\times 10^{-22}\ecm$. 

{\bf Extra dimensional and composite models}
\newline
In models with an extra dimension (such as the Randall--Sundrum model~\cite{Randall:1999ee}) or models with a strongly coupled Higgs sector (e.g.\ the littlest Higgs model~\cite{ArkaniHamed:2002qy}) one obtains in general new massive fermions and vectors that are resonances of the SM particles~\cite{Contino:2006nn}. Assuming a new quantum number called $T$-parity~\cite{Cheng:2003ju}, these particles do not mix with the SM and one obtains a scenario as above with heavy new fermions and vector bosons as in~\eqref{Lagr_vector}. However, it has been shown in~\cite{Blanke:2007db} that for the littlest Higgs model with $T$-parity the effect in $a_\mu$ is small since the couplings of the resonances are mainly left-handed and therefore do not allow for a sufficient chiral enhancement. The same is true for generic RS models~\cite{Perez:2008ee,Beneke:2012ie}. Again, explaining $a_\mu$ and $a_e$ in the simplest models is not possible since the vector resonances are not flavor specific and have common couplings to muons and electrons, violating the $\mu\to e\gamma$ bound.  

{\bf MSSM (with large $\boldsymbol{A}$ terms)}
\\
$a_\mu$ in the MSSM has been most extensively studied in the context of the constrained MSSM or with the assumption of flavor-universal supersymmetry (SUSY) breaking terms (see e.g.~\cite{Stockinger:2006zn} for a review), 
i.e.\ respecting (naive) MFV. As outlined in the introduction, a model with MFV cannot explain $a_\mu$ and $a_e$ simultaneously, and this is of course also true for the MSSM. Furthermore, the phase in $c_R^{\mu\mu}$ is correlated with $c_R^{ee}$ suppressing possible effects in the muon EDM.

However, since the MSSM possesses three generations of sleptons, effects in electrons and muons can in principle be decoupled. In fact, with a general flavor structure of the SUSY breaking terms, contributions of large non-universal trilinear $A$-terms can in principle give the right effects~\cite{Borzumati:1999sp,Crivellin:2010ty}. Still, large $A^\mu$ and, even more significantly, large $A^e$-terms are delicate because of fine-tuning in the lepton mass matrix~\cite{Crivellin:2010gw} and charge-breaking minima of the scalar potential that would render the vacuum unstable~\cite{Gunion:1987qv}. The latter constraints could be avoided by using non-holomorphic $A^\prime$ terms~\cite{Haber:2007dj}. Nonetheless, achieving such anarchic $A$-terms, while still respecting all other flavor bounds, with a SUSY breaking mechanism seems very challenging.\footnote{Alternatively, one can also use flavor-violating SUSY breaking terms together with chirality violation from the $\tau$ sector to generate an enhanced effect 
in $a_\mu$ that has a free phase and can therefore also generate a large effect in $d_\mu$~\cite{Hiller:2010ib}. However, in this case more free parameters are involved and constraints from $\tau\to\mu\gamma$ arise.}

\subsection{Model with a new scalar and fermions}

Due to the shortcomings of the models discussed so far in providing a common explanation of $a_\mu$ and $a_e$ we turn now to models in which we introduce vector-like generations of leptons.
The general comments regarding the chiral enhancement still apply, so special attention will be paid to the flavor structure of the corresponding models.

\begin{figure}
\centering
\includegraphics[height=\linewidth,angle=-90]{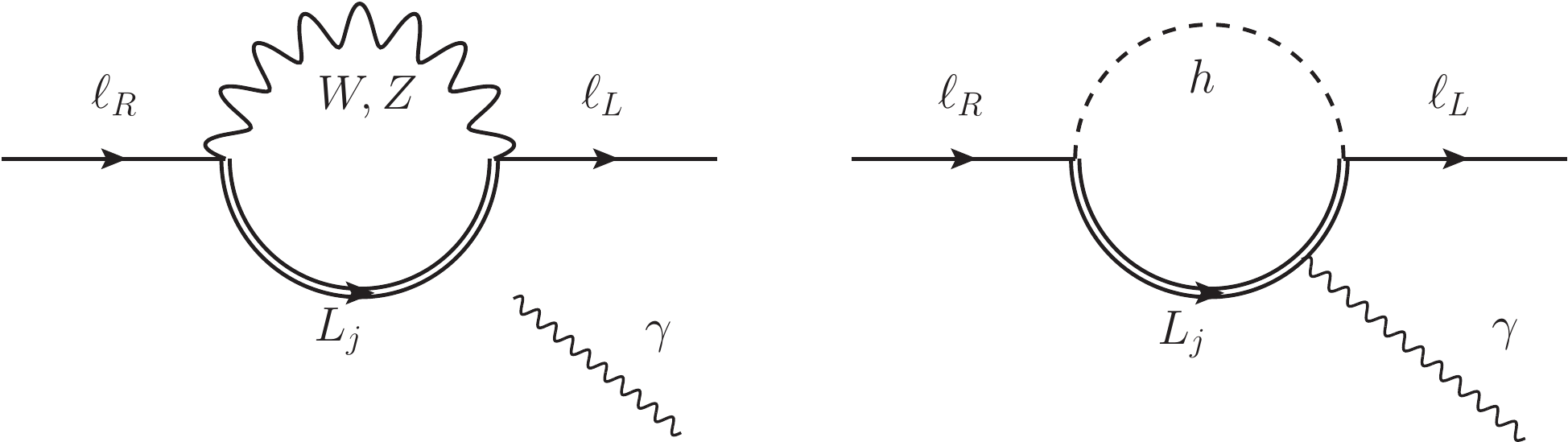}
\caption{Generic diagrams contributing to the dipole operator in Model I.}
\label{fig:1loop}
\end{figure}

{\bf Model I}
\newline
Let us introduce three $SU(2)$ doublets $L_i$ and three $SU(2)$ singlets $E_i$ with the same quantum number as the SM lepton doublet and singlet, respectively.\footnote{$a_\mu$ in models with one generation of vector-like leptons has been studied in~\cite{Czarnecki:2001pv,Kannike:2011ng,Dermisek:2013gta,Freitas:2014pua,Aboubrahim:2016xuz,Kowalska:2017iqv,Raby:2017igl,Calibbi:2018rzv}.} In order to avoid the bound from $\mu\to e\gamma$ we assume that these three generations of heavy leptons couple separately to muons and electrons (and taus). This flavor conservation can be guaranteed by introducing an Abelian flavor symmetry 
for the leptons and their vector-like partners, e.g.\ $L_\mu-L_\tau$~\cite{He:1990pn,Foot:1990mn}, but also other charge assignments are compatible with the observed PMNS matrix~\cite{Araki:2012ip} and ensure flavor conservations as well. 
Therefore, it is sufficient to discuss the case for each charged fermion $\ell=e,\mu,\tau$ separately.
The generic diagrams to be considered for $c_R$ are shown in Fig.~\ref{fig:1loop}.

We start from the following Lagrangians for the mass terms and the interactions with the Higgs of the vector-like leptons 
\begin{align}
\Lagr_M&=  - {M_L}{{\bar L}_L}{L_R} - {M_E}{{\bar E}_L}{E_R} + \text{h.c.},\notag\\
\Lagr_H&=  - {\kappa _L}{{\bar L}_L}H{E_R} - \kappa _E^{}{{\bar L}_R}H{E_L} \notag\\
&- {\lambda _L}{{\bar L}_L}{\ell_R}H - {\lambda _E}{{\bar E}_R}\tilde H{\ell _L} + \text{h.c.}
\end{align}
In principle, there are also mass terms connecting SM leptons to their heavy partners. However, in our setup these terms can always be absorbed into a redefinition of the fields. 

This model gives rise to tree-level effects in the $Z\to\ell\ell$ and $h\to\ell\ell$ couplings. At leading order in $v/M$ the corrections to $Z\to\ell\ell$ are given by
\begin{align}
\Lagr_Z&=\sum_{i=L,R}\big(Z_{\ell\ell}^i+\Delta Z_{\ell\ell}^i\big)\bar \ell \gamma^\mu P_i \ell Z_\mu,\notag\\
\Delta Z_{\ell \ell}^L&=\frac{v^2|\lambda _E|^2}{M_E^2}\big(Z_{\ell \ell}^R-Z_{\ell \ell }^L\big),\notag\\
\Delta Z_{\ell \ell}^R&= \frac{v^2|\lambda _L|^2}{M_L^2}\big(Z_{\ell \ell }^L-Z_{\ell \ell }^R\big),
\end{align}
with $g=\sqrt{2}M_W/v$,
\begin{align}
Z_{\ell \ell }^L &= \frac{g}{2\cos\theta_W}\big(1 - 2\sin^2\theta_W\big),\notag\\
Z_{\ell \ell }^R &=  - \frac{g}{c_W}\sin^2\theta_W,
\end{align}
 and the Higgs vacuum expectation value $v\sim 174\GeV$ ($H^0= v+ h/\sqrt{2}$). Similar corrections also pertain to the $W$ couplings
\beq
\Delta W_{\ell \ell}^L=-\frac{v^2|\lambda _E|^2}{2M_E^2}W^L_{\ell\ell}\equiv\delta W^L_{\ell\ell}W^L_{\ell\ell},
\eeq
but since they are less well constrained we only consider the indirect impact on $Z\to\ell\ell$ 
due to the renormalization of $G_F$ extracted from muon decay 
\beq
\Delta Z_{\ell \ell }^{L,R}\to \Delta Z_{\ell \ell }^{L,R} -\frac{1}{2}Z_{\ell \ell }^{L,R}\big( \delta W^L_{\mu\mu}+\delta W^L_{ee}\big).
\eeq

For $h\to\ell\ell$ we find a shift of the Yukawa coupling (normalized to the one of the SM) given by
\beq
\delta Y^\ell = \frac{\Delta Y^\ell}{Y^\ell} = 2\kappa_E^*\lambda _E^*{\lambda _L}\frac{v^3}{m_\ell M_LM_E}.
\eeq
The heavy fermions $E$, (second component of) $L$, and the charged SM lepton mix to mass eigenstates $\chi^{\pm}$~\cite{Dermisek:2013gta}, so that 
after diagonalization of the mass matrix we derive the contribution to dipole moments from the general relation~\eqref{c_R_general_Phi} with $Q=-1$
\begin{align}
\label{c_R_model_I_Higgs}
c_R^{\text{I}, h}&= - \frac{e}{32\pi^2}\frac{v\lambda _E^*\lambda _L \kappa_E^*\Big[M_E^2 F_\Phi\Big(\frac{M_E^2}{m_h^2}\Big) + M_L^2 F_\Phi\Big(\frac{M_L^2}{m_h^2}\Big)\Big]}{m_h^2M_L M_E}\notag\\
&+\frac{e}{32\pi^2}\frac{v\lambda _E^*\lambda _L}{m_h^2(M_E^2-M_L^2)}\notag\\
&\times\bigg[M_E\big(M_E\kappa _L^* + M_L\kappa _E^*\big)F_\Phi\bigg(\frac{M_E^2}{m_h^2}\bigg)\notag\\
&- M_L\big(M_E\kappa _E^* + M_L\kappa _L^* \big)F_\Phi\bigg(\frac{M_L^2}{m_h^2} \bigg)\bigg],
\end{align}
where $F_\Phi(x)=f_\Phi(x)-g_\Phi(x)$ and we have again expanded in $v/M$. In the limit $m_h\ll M_E=M_L=M$ this expression simplifies to
\beq
\label{c_R_asym_Higgs}
c_R^{\text{I}, h}=\frac{e}{16\pi^2} \frac{3v\lambda _E^*\lambda _L \kappa _E^*}{8M^2}.
\eeq
In practice, the pieces that are not chirally enhanced can indeed be safely ignored, and were already dropped in~\eqref{c_R_model_I_Higgs}, but, in particular for low masses, higher orders in $v$ can become relevant. Accordingly, we keep the exact diagonalization everywhere in the numerical analysis, including the $Z\to\ell\ell$ couplings. 

Moreover, in the asymptotic $M\to\infty$ limit the Higgs contribution is dominant, but in the general case we do need to keep the $Z$ and $W$ loops
\begin{align}
 c_R^{\text{I}, Z}&= -\frac{e}{32\pi^2}\frac{v\lambda _E^*\lambda _L}{(M_E^2-M_L^2)M_E M_L}\notag\\
 &\times \bigg[M_E\big(M_E\kappa _E^* + M_L\kappa _L^*\big)F_V\bigg(\frac{M_E^2}{M_Z^2}\bigg)\notag\\
 &-M_L\big(M_L\kappa_E^* + M_E\kappa _L^*\big)F_V\bigg(\frac{M_L^2}{M_Z^2}\bigg)\bigg],\notag\\
 c_R^{\text{I}, W}&=-\frac{e}{16\pi^2}\frac{v\lambda _E^*\lambda _L \kappa_E^*}{M_E M_L}f_V\bigg(\frac{M_L^2}{M_W^2}\bigg),
\end{align}
with $F_V(x)=f_V(x)-g_V(x)$, as derived from~\eqref{c_R_general_V} with $Q=-1$ and $Q=0$, respectively. Asymptotically, they approach 
\beq
c_R^{\text{I}, Z}=\frac{e}{16\pi^2} \frac{v\lambda _E^*\lambda _L \kappa _E^*}{8M^2},\qquad
c_R^{\text{I}, W}=-\frac{e}{16\pi^2} \frac{v\lambda _E^*\lambda _L \kappa _E^*}{4M^2},
\eeq
and are thus suppressed by factors $3$ and $-3/2$ compared to the Higgs contribution. For smaller masses, however, they become dominant, especially the $W$ loop.

\begin{figure*}
	\centering
		\includegraphics[height=7cm]{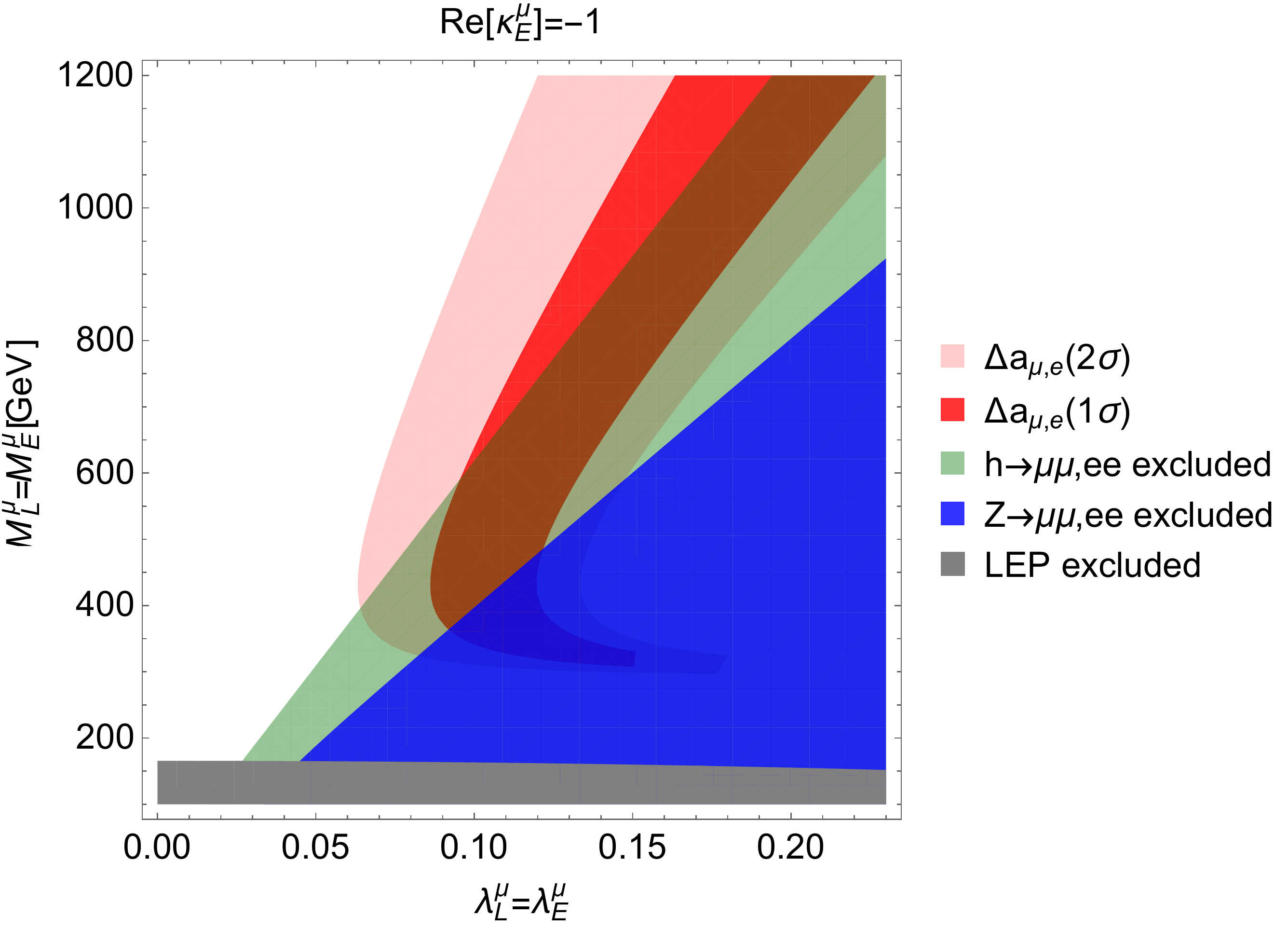}
		\includegraphics[height=7cm]{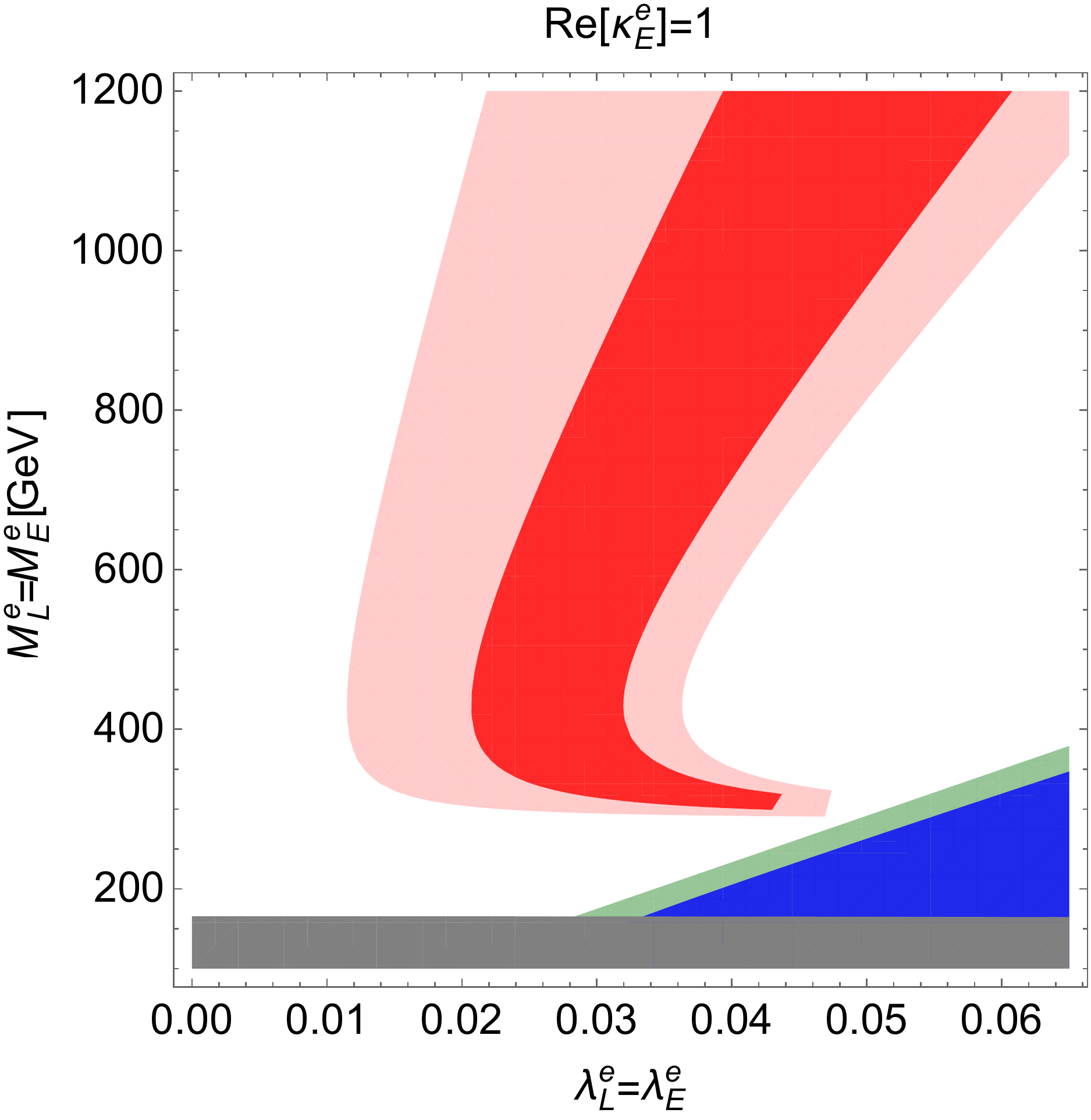}
	\caption{Allowed regions of $a_\ell$ in the $\lambda_E=\lambda_L$--$M_E=M_L$ plane for $\kappa_L=0$ and $\kappa_E=\mp1$ for muon (left) and electron (right).
	The bounds are derived from $\sigma(h\to\mu^+\mu^-)/\sigma(h\to\mu^+\mu^-)_\text{SM}=0\pm 1.3$~\cite{Patrignani:2016xqp,Khachatryan:2016vau,Aaboud:2017ojs},
	$\sigma(h\to e^+e^-)/\sigma(h\to e^+e^-)_\text{SM}<3.7\times 10^5$~\cite{Khachatryan:2014aep}, $Z\to\ell\ell$~\cite{Patrignani:2016xqp,ALEPH:2005ab},
	and direct searches for new heavy charged leptons~\cite{Achard:2001qw}. The $h\to\ell\ell$ limits are implemented at $2\sigma$, the ones for $Z\to\ell\ell$ at $3\sigma$, as explained in the main text.}
	\label{aellHiggs}
\end{figure*}

The numerical results in the limit $\lambda_E=\lambda_L$ and $M_E=M_L$ are shown in Fig.~\ref{aellHiggs}, for $a_\mu$ (left) and $a_e$ (right). 
The restrictions from the modified $Z\to\ell\ell$ couplings that arise indirectly from the opposite channel via muon decay prove to be weaker than the direct ones and are therefore neglected, 
while the corrections from muon decay within the same channel are kept. In general, we implement all constraints at the $2\sigma$ level, 
the exception being the $Z\to \ell\ell$ couplings for which we allow for a $3\sigma$ ellipse. 
This treatment is motivated by the fact that there is a tension between the measured $Z\to ee$ couplings and their SM values at the level of $2\sigma$, so that a slightly higher significance is required to ensure
that the SM is included in the allowed range.

We find that $\Delta a_\mu$ can only be explained for relatively large couplings, consequently with a parameter space already partly excluded by $h\to\mu\mu$ and $Z\to\mu\mu$. 
In contrast, $\Delta a_e$ is still allowed for a wide range of masses and couplings. 
The reason for this behavior can be understood already from the relative size of the respective deviation, see~\eqref{size}, 
in such a way that all other parameters being equal $\Delta a_\mu$ requires larger couplings.
The different sign in $\kappa_E$ reflects the respective sign in $\Delta a_\ell$ and due to the
dominance of the Higgs contribution for asymptotic masses can be read off from~\eqref{c_R_asym_Higgs}.
We conclude that while for the muon the construction based on the SM Higgs is already under pressure, $\Delta a_e$ does 
permit such an explanation for a wide range of parameters. However, a huge relative effect in the effective electron Yukawa coupling (which could be considered as fine-tuning) appears, 
enhancing $h\to ee$ by orders of magnitude compared to the SM with potentially interesting phenomenological consequences~\cite{Altmannshofer:2015qra}.

{\bf Model II}
\newline
One way to avoid the $Z\to\ell\ell$ and $h\to\ell\ell$ bounds discussed above is to introduce a new scalar $\phi$ charged under a $Z_2$ symmetry, which takes the role of the SM Higgs. 
In this way, only the interaction term
\beq
\Lagr_\phi=-\lambda_L^\phi \bar L_R\phi \ell_L-\lambda_E^\phi \bar E_L \phi \ell_R+\text{h.c.}
\eeq
changes.
This SM extension leads to chirally enhanced effects in $a_\mu$ and $a_e$, while the $Z_2$ symmetry forbids $\bar L_L H e_R$ couplings that would otherwise produce the potentially problematic 
couplings. In analogy to~\eqref{c_R_model_I_Higgs} we obtain
\begin{align}
c_R^\text{II}&= \frac{e}{16\pi^2}\frac{v\lambda _E^{\phi}\lambda_L^{\phi*}}{m_\phi^2(M_E^2-M_L^2)}\notag\\
&\times\bigg[M_E\big(M_E\kappa_E + M_L\kappa_L\big)F_\Phi\bigg(\frac{M_E^2}{m_\phi^2}\bigg)\notag\\
&- M_L\big(M_E\kappa_L + M_L\kappa_E \big)F_\Phi\bigg(\frac{M_L^2}{m_\phi^2} \bigg)\bigg],
\end{align}
which in the limit $m_\phi=M_E=M_L=M$ simplifies to
\beq
\label{cR_simp}
c_R^\text{II} =-\frac{e}{16\pi^2}\frac{v \lambda}{24M^2},\qquad \lambda=\lambda _E^{\phi}\lambda_L^{\phi*}(\kappa_E-3\kappa_L).
\eeq
The resulting expression 
\beq
a_\ell = \lambda_\ell \frac{v m_\ell}{96\pi^2 M^2} 
\eeq
can then easily explain both anomalies, for $M=1\TeV$ with $\lambda_e\sim -0.01$ and $\lambda_\mu\sim 0.15$, but thanks to the chiral enhancement the original couplings 
$\lambda_{E,L}^\phi$ and $\kappa_{E,L}$ remain perturbative up to at least $M=10\TeV$, again at the price of decoupling the $\mu$ and $e$ sectors. 

With slight modification of the model it might be possible to address other issues that require BSM input. For instance, we can add a charged scalar instead of a neutral one by changing the hypercharge of the new (Majorana) fermion fields accordingly. In this way, the two neutral components would mix to mass eigenstates $\chi^0$, which could produce a dark matter candidate. Moreover, these states could 
generate neutrino masses by the type I see-saw mechanism, and, assuming a non-vanishing phase in $\lambda_\ell$, produce a $CP$ asymmetry that could be relevant for creating 
the matter--anti-matter asymmetry via leptogenesis. Such a phase, in turn, could be observable as a larger-than-MFV muon EDM, see Sect.~\ref{sec:muEDM}.   

{\bf Model III}
\newline
Interestingly, one can combine the two effects of models I and II for explaining $a_\mu$ and $a_e$ to actually decrease the particle content. To this end, one can explain $a_e$ via an effect 
induced by the SM Higgs (and also modified $Z$ and $W$ couplings) while one accounts for $a_\mu$ through an additional neutral scalar. In this setup, it is sufficient to add a single vector-like generation. Assuming $L_\mu-L_\tau$ as the flavor symmetry, the vector-like generation is uncharged (coupling to electrons and the SM Higgs) while the additional scalar is singly charged (coupling to 
the vector-like generation and muons). Accordingly, the Lagrangian for scalar and Higgs interactions of this model takes the form
\begin{align}
\Lagr&= -\lambda_L^\phi \bar L_R\phi \mu_L-\lambda_E^\phi \bar E_L \phi \mu_R \notag\\
&-\lambda _L{{\bar L}_L}{e_R}H - {\lambda _E}{{\bar E}_R}\tilde H{e _L} + \text{h.c.}
\end{align}
Furthermore, once the flavor symmetry is gauged, one can identify the neutral scalar with the ``flavon,'' and the scenario becomes very similar to the model of~\cite{Altmannshofer:2016oaq}.\footnote{Since the hint for an excess in $h\to\tau\mu$ disappeared, we do not need the second scalar introduced in~\cite{Altmannshofer:2016oaq} and the flavon can be heavy.} Moreover, gauging $L_\mu-L_\tau$ allows for explaining the intriguing hints for lepton flavor non-universality in $b\to s\ell\ell$ transitions (which are currently at the $5\sigma$ level~\cite{Capdevila:2017bsm}) once vector-like quarks are added~\cite{Altmannshofer:2014cfa,Crivellin:2015mga,Bobeth:2016llm,Ko:2017yrd}, quarks are charged under the same flavor symmetry~\cite{Crivellin:2015lwa}, or mix with another $Z^\prime$~\cite{Crivellin:2016ejn}. However, once the flavon acquires a vacuum expectation value, again non-zero rates for $\mu\to e\gamma$ arise and a more intricate structure would be needed to sufficiently suppress it.

\section{A large muon EDM}
\label{sec:muEDM}

All the examples discussed in the previous section have in common that agreement with $\Delta a_\mu$ and $\Delta a_e$, if taken at face value, demands the decoupling of the $\mu$ and $e$ BSM sectors due to the constraints from $\mu\to e\gamma$. In such a scenario, the stringent limits on the electron EDM~\eqref{delimit} do not constrain the phase of $c_R^{\mu\mu}$,
which, as argued above, leads one to expect $|d_\mu|=\Order(10^{-22}\ecm)$. Accordingly, the best constraint on the phase of $c_R^{\mu\mu}$, derived from the present limit on the muon EDM, only excludes values very close to $90^{\circ}$, see~\eqref{limits_phase}, and leaves open the possibility of a muon EDM much larger than expected from MFV scaling~\eqref{dmu_MFV}.

\begin{figure}
\centering
\includegraphics[width=0.5\linewidth]{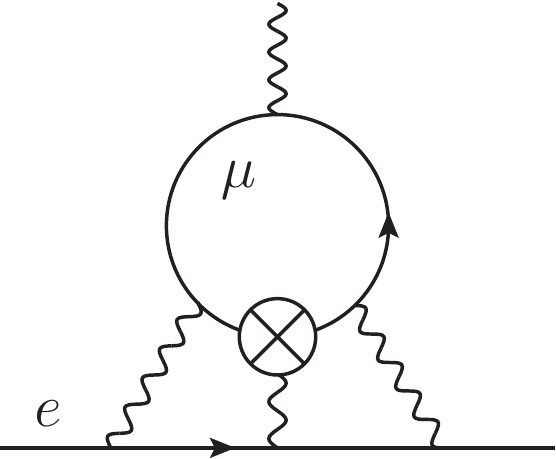}
\caption{Three-loop diagram that produces a contribution to the electron EDM by an insertion of the muon EDM operator indicated by the cross. The other diagrams with insertions 
at the remaining muon--photon vertices as well as the permutations at the electron line are not shown.}
\label{fig:3loop}
\end{figure}

\begin{figure}
\centering
\includegraphics[width=\linewidth]{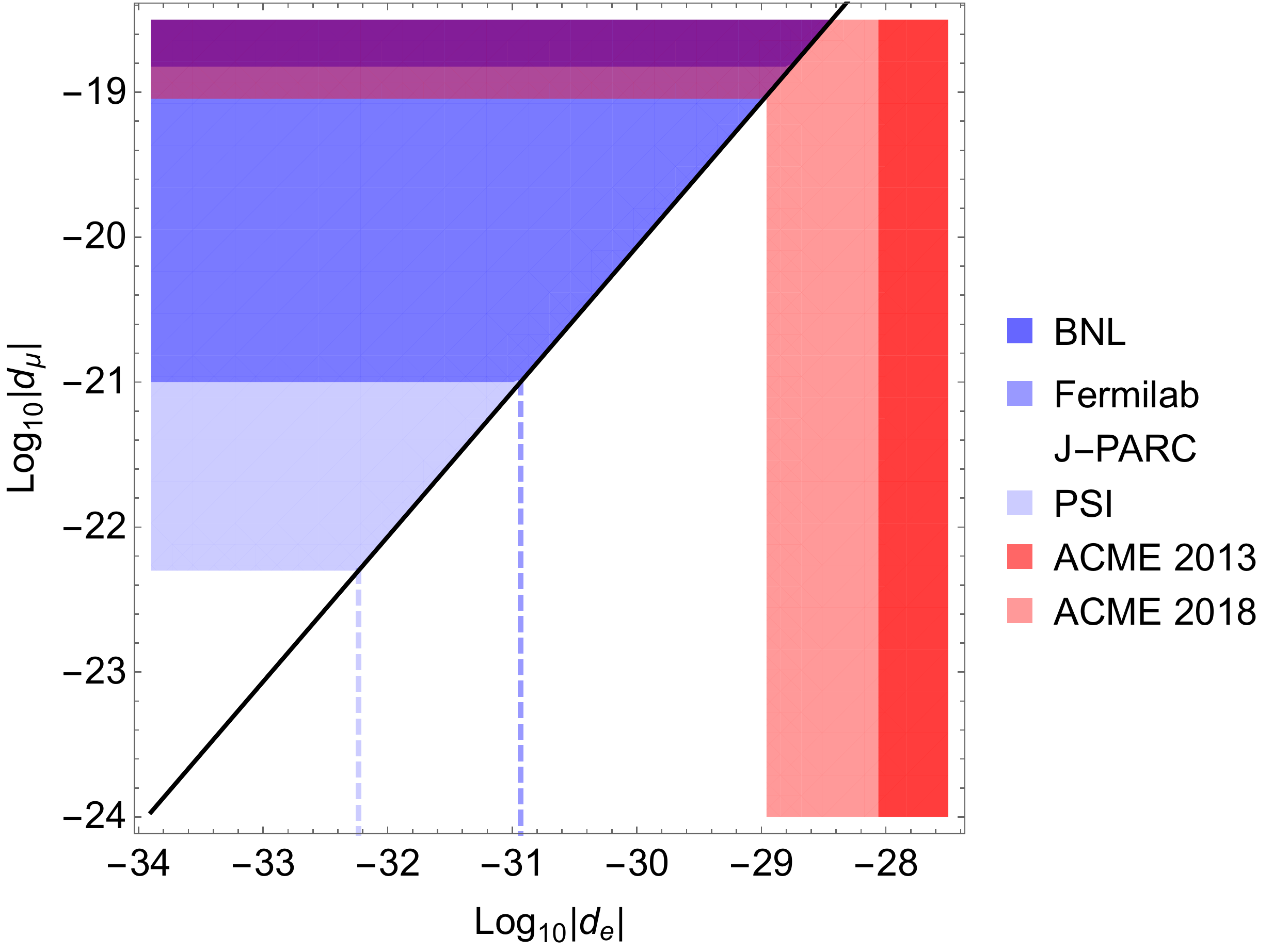}
\caption{Present and future direct limits on $|d_\mu|$ from BNL~\cite{Bennett:2008dy} (dark blue), see~\eqref{dmulimit}, Fermilab/J-PARC (blue), and the proposed PSI experiment (light blue). The dark red and light red regions refer to the ACME 2013~\cite{Baron:2013eja} and ACME 2018~\cite{Andreev:2018ayy} limits on $|d_e|$, respectively, where the latter provides an indirect bound on $|d_\mu|$ slightly 
better than the BNL direct bound. The blue dashed lines indicate the limits on $|d_e|$ that would be required to match the anticipated direct limits from Fermilab/J-PARC and PSI.
The black line defines the relation~\eqref{indirect_bound}, with the upper-left half referring to limits on $|d_\mu|$ and the lower-right to limits on $|d_e|$.}
\label{fig:dedmu}
\end{figure}

However, despite the absence of a direct correlation, indirect bounds on $|d_\mu|$ can still be extracted from limits on $|d_e|$ by means of the three-loop diagram shown in Fig.~\ref{fig:3loop}, i.e.\ 
from the contribution of the muon dipole operator to the electron EDM (see~\cite{Grozin:2009jq}, where this argument was used to derive improved limits on the EDM of the $\tau$). This indirect limit produces
\begin{align}
\label{indirect_bound}
|d_\mu|&\leq \Bigg[\bigg(\frac{15}{4}\zeta(3)-\frac{31}{12}\bigg)\frac{m_e}{m_\mu}\bigg(\frac{\alpha}{\pi}\bigg)^3\Bigg]^{-1}|d_e|\notag\\
&\leq 0.9\times 10^{-19}\ecm\qquad 90\% \,\text{C.L.},
\end{align}
slightly better than the direct limit~\eqref{dmulimit}. Therefore, the limit on $|d_e|$, which is $10$ orders of magnitude better than the limit on $|d_\mu|$, just barely suffices to 
overcome the three-loop suppression, but it is unlikely that the result can be further improved by orders of magnitude, see Fig.~\ref{fig:dedmu}. 
For this reason we now turn to the prospects of improving the direct limit on the muon EDM.

\begin{figure*}%
	\centering
	\includegraphics[width=0.47\linewidth]{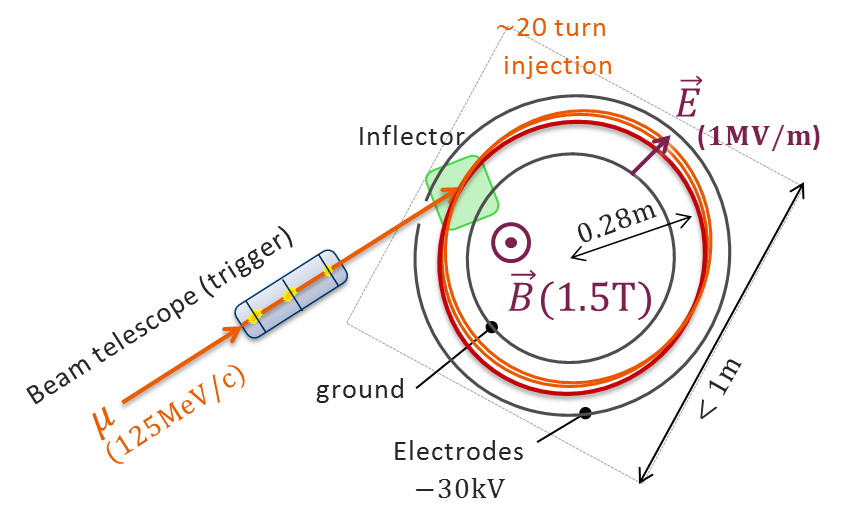}%
	\hfill
	\includegraphics[width=0.4\linewidth]{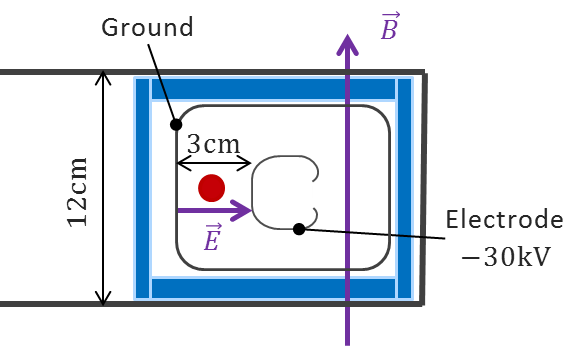}
	\caption{Sketch (a) and cross section (b) of the compact storage ring setup to search for a muon EDM (not to scale). (a) Polarized muons ($P\approx0.9$) from in-flight decays of pions with a momentum 
		of $200$\MeV/c from target E (not shown) of the high intensity proton accelerator of PSI arrive every $19.75$\,\text{ns} at the beam telescope. A dipole magnet with  $B=1.5\,\text{T}$ with a circular electrode system forms the magnetic storage ring. Once the beam telescope identifies a muon within the acceptance of the storage ring, the inflector, an air coil perturbing the otherwise homogeneous field, will be synchronously ramped down using a 1/2 integer resonant injection scheme. After about twenty turns (orange orbits), the muon will stay on the stable (red) orbit until it decays. 
		A positron tracker around the orbit will count the decay positrons relative to time. In general only one muon at a time will be stored. A veto sends the muon beam on a beam dump (not shown) until the decay is registered or a time-out occurs. Then the next cycle starts with ramping of the inflector field. (b) The muon orbit (red spot) is surrounded by a positron tracker system (blue) inside a vacuum chamber (gray outer lines). An  electrode system encapsulated in a ground ring creates the electric field.}%
	\label{fig:muEDMSketch}%
\end{figure*}

\section{Experimental prospects}

The first search for the muon EDM resulted in an upper limit of $2.9\times 10^{-15}\ecm$ (95\% C.L.)~\cite{Berley:1958} and was published in 1958. Half a century later the current best upper limit~\eqref{dmulimit} was deduced using the spin precession data from the $(g-2)_\mu$ storage ring experiment E821 at BNL~\cite{Bennett:2006fi}. 

The EDM can be similarly defined as the magnetic moment $\boldsymbol{\mu}=gq\hbar\boldsymbol{\sigma}/(4mc)$, leading to
\begin{equation} 
\boldsymbol{d} = \eta\frac{q\hbar}{4mc}\boldsymbol{\sigma},
\label{eq:LeptonEDM}
\end{equation}
where $q$, $m$, $\boldsymbol{\sigma}$ are the elementary charge, mass, and spin of the muon. 
Hence, the spin precession $\boldsymbol{\omega}$ of a muon in a storage ring with an electric field $\boldsymbol{E}$ and magnetic field $\boldsymbol{B}$ is given by
\begin{align}
	\boldsymbol{\omega}&=\frac{q}{m}\left[a\boldsymbol{B}-\left(a+\frac{1}{1-\gamma^2}\right)
	\frac{\boldsymbol{\beta}\times\boldsymbol{E}}{c}\right]\notag\\
	&+\frac{q}{m}\frac{\eta}{2}\left(\boldsymbol{\beta}\times\boldsymbol{B}+\frac{\boldsymbol{E}}{c}\right),
\label{eq:omegaMu1}
\end{align}
where $a=(g-2)/2$ is the anomalous magnetic moment and $\gamma=1/\sqrt{1-\beta^2}$. The first term in~\eqref{eq:omegaMu1} is the anomalous precession frequency $\boldsymbol{\omega}_\text{a}$, 
the difference of the Larmor precession and the cyclotron precession oriented parallel to the magnetic field. The second term is the precession $\boldsymbol{\omega}_\text{e}$ due to an EDM 
coupling to the relativistic electric field of the muon moving in the magnetic field $\boldsymbol{B}$, oriented perpendicular to $\boldsymbol{B}$.
In the case of the E821 experiment muons with a so called ``magic'' momentum of $p_\text{magic} = m/\sqrt{a} = 3.09\GeV/c$ were used, simplifying~\eqref{eq:omegaMu1} to
\begin{equation}
	\boldsymbol{\omega}=\frac{q}{m}\left[a\boldsymbol{B}+\frac{\eta}{2}\left(\boldsymbol{\beta}\times\boldsymbol{B}+\frac{\boldsymbol{E}}{c}\right)\right],
\label{eq:MagicOmega}
\end{equation}
which makes the anomalous precession frequency independent of electric fields needed for steering the beam. 
In this case the precession plane is tilted out of the orbital plane defined by the movement of the muon in the presence of an EDM. 
Hence, a vertical precession ($\boldsymbol{\omega}_\text{e} \bot \boldsymbol{B}$) with an amplitude proportional to the EDM with a frequency 
$\omega$ phase-shifted by $90^{\circ}$ with respect to the horizontal anomalous precession becomes observable. 
Another effect of an EDM is the increase of the observed precession frequency 
\begin{equation}
\omega=\sqrt{\boldsymbol{\omega}_\text{a}^2 + \boldsymbol{\omega}_\text{e}^2}.
\label{eq:SumOfPrecession}
\end{equation}
Three different data sets from different detectors of the BNL E821 experiments were used to search for a muon EDM signal~\cite{Bennett:2008dy}, which resulted in the current best measurement of 
$d_\mu = -0.1(9)\times 10^{-19}\ecm$.
In the meantime, the Muon $g-2$ collaboration moved the storage ring to Fermilab and upgraded the detection system. A first data-taking campaign was just completed. 
More statistics and reduced systematic shall eventually lead to a new muon EDM search with a sensitivity down to $10^{-21}\ecm$~\cite{Chislett:2016jau},
and a similar sensitivity could be expected from the J-PARC $(g-2)_\mu$ experiment~\cite{Gorringe:2015cma}. 

A further increase of the sensitivity for a muon EDM search is possible employing the frozen-spin technique~\cite{Semertzidis:1999kv,Farley:2003wt}. This requires tuning of the 
electric and magnetic fields in~\eqref{eq:omegaMu1} in such a way that the first term cancels
\begin{equation}
	a\boldsymbol{B}-\left(a+\frac{1}{1-\gamma^2}\right)
	\frac{\boldsymbol{\beta}\times\boldsymbol{E}}{c} = 0.
\label{eq:FrozenSpinCondition}
\end{equation}
In this case with $\eta =0$, either absent or a negligibly small muon EDM, the spin exactly follows the momentum, and in the rest frame of the muon the spin is ``frozen.''

\subsection{Prospects of the frozen-spin technique using a compact storage ring at PSI}

In~\cite{Adelmann:2010zz} it has been shown that a dedicated compact muon storage ring at PSI employing the frozen-spin technique is an attractive method to search for a muon EDM. 
The proposed design uses positively charged muons with a momentum of $125$\MeV/c, corresponding to a velocity of $\beta c =0.766c\approx 23\,\text{cm}/\text{ns}$ from the muon beam line $\mu$E1 at PSI, in combination with a fast trigger system based on a muon telescope and a storage ring made of a very homogeneous conventional dipole magnet with a field of $B=1.5\,\text{T}$. A sketch of the experiment is shown in Fig.~\ref{fig:muEDMSketch} (a), while a cross section of the orbit with positron detectors and electric field electrodes is shown in Fig.~\ref{fig:muEDMSketch} (b).

In order to maximize the sensitivity to an EDM of the muon, we will employ the frozen-spin technique with 
\begin{equation}
E\approx a B c\beta\gamma^2,
\label{eq:FrozenSpin}
\end{equation}
which eliminates the anomalous precession signal. In the case of an EDM ($\eta\neq 0$) the spin will start to precess out of the orbital plane. A positron-detection system around the storage ring will 
detect the decay positrons. Due to the average decay asymmetry $\bar\alpha$, more positrons are emitted along the muon spin. This will lead to a build-up of an up/down asymmetry with time, 
proportional to $\eta$, the EDM signal.

For the frozen-spin technique the sensitivity is given by, see Eq.~(5) of~\cite{Adelmann:2010zz},
\begin{equation}
	\sigma(d_{\mu}) =\frac{ \hbar \gamma a}{2\tau E \bar\alpha P \sqrt{N}}
\label{eq:sensitivity}
\end{equation}
for a polarization $P$, the muon life time $\tau=2.2\,\mu\text{s}$, and the number of detected positrons $N$ (we used~\eqref{eq:LeptonEDM} to replace $\eta$). 
In a magnetic field of $B=1.5\,\text{T}$ a radial electric field of $E=1\,\text{MV}/\text{m}$ is required for the frozen-spin technique, resulting in a radius 
$r=0.28\,\text{m}$. With $\bar{\alpha}=0.3$, $P=0.9$, and $N=4\times 10^{14}$ per year, one expects a statistical sensitivity of
\begin{equation}
	\sigma(d_{\mu}) = 5\times 10^{-23}\ecm,
\label{eq:ExpectedSensitiviy1Year}
\end{equation}
for one year of data taking. 
A possible method to further increase the experimental sensitivity of the experiment is to use a higher muon momentum up to $200\MeV/c$~\cite{Petitjean:2018}, 
which will be studied in detail using simulations. Assuming the muon polarization remains the same, this would result in a twofold improvement of sensitivity requiring higher electric fields 
of $E=22\,\text{kV}/\text{cm}$ and a radius of $r=0.44\,\text{m}$.

\begin{figure}
	\centering
	\includegraphics[width=0.8\linewidth]{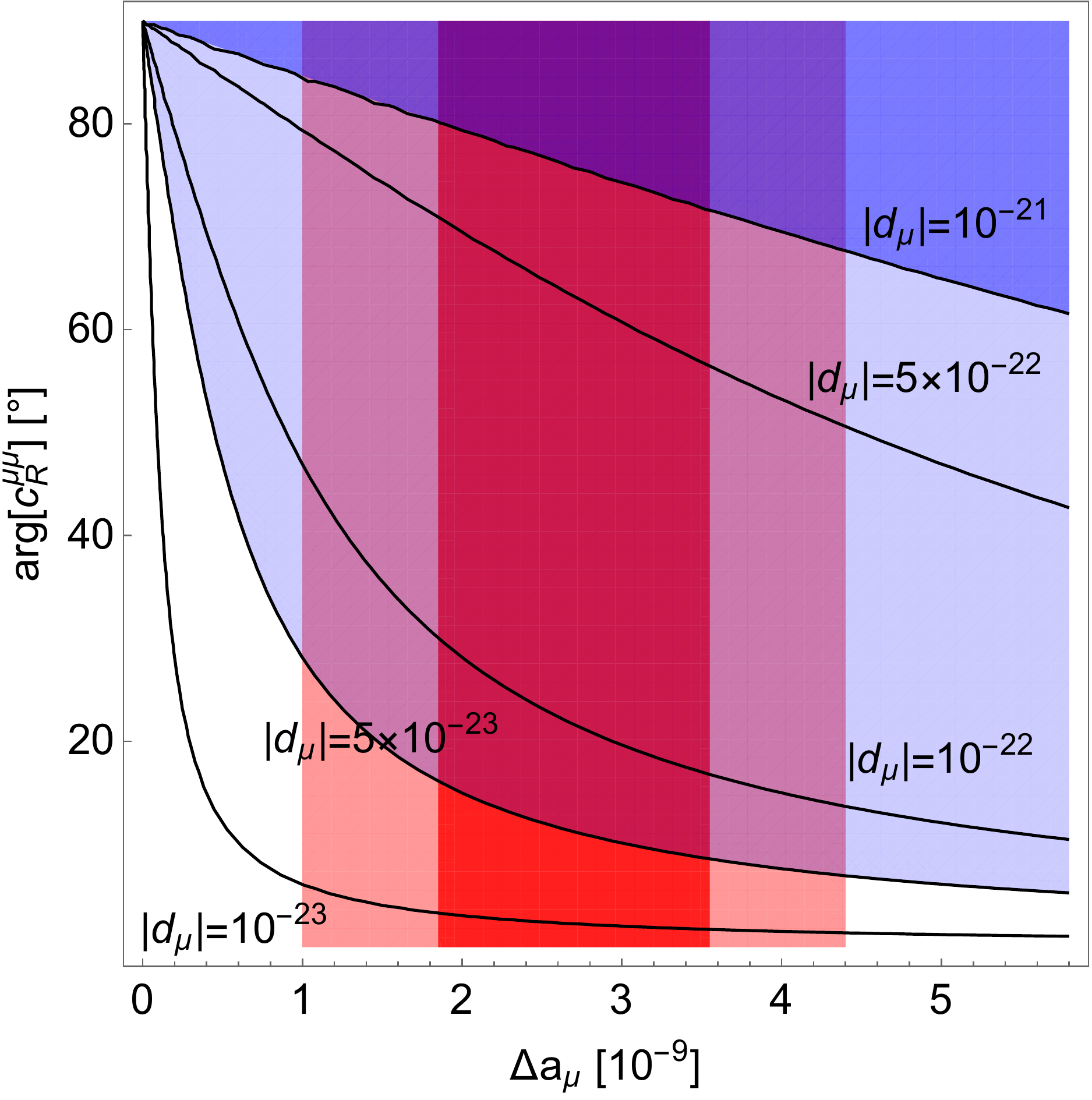}
	\caption{Contour lines defining the muon EDM (in units of $\ecm$) as a function of $\Delta a_\mu$ and the phase of the Wilson coefficient $c_R^{\mu\mu}$. The red regions are currently preferred by the 
		measurement of $a_\mu$ and the blue regions are the expected sensitivity of the Fermilab/J-PARC (dark blue) and the proposed PSI experiment (light blue). Note that since a chirally 
		enhanced effect is preferred, $\text{arg}[c_R^{\mu\mu}]$ is a free phase of the theory and in general expected to fulfill $\tan(\text{arg}[c_R^{\mu\mu}])=\Order(1)$. The limit on the phase derived from the current limit 
		for $|d_\mu|$ is so close to $90^{\circ}$ that it is not visible in the plot.}
	\label{fig:sensitivity}
\end{figure}

The experimental prospects for the muon EDM are summarized in Fig.~\ref{fig:sensitivity}. Depending on the value of $a_\mu$, 
the current limit is too weak to impose a visible bound on the phase of $c_R^{\mu\mu}$. While the future $(g-2)_\mu$ experiments at Fermilab and J-PARC can only cover phases above approximately $70^{\circ}$, 
the proposed frozen-spin technique experiment at PSI would be sensitive down to small phases of around $10^{\circ}$.

\section{Conclusions}

In this article we argued that the recent tension observed in $a_e$, together with the long-standing anomaly in $a_\mu$,  can be considered an indication that potential BSM physics does not respect MFV. Furthermore, due to the constraint from $\mu\to e\gamma$, the electron and muon sectors need to be (nearly) completely decoupled from each other. We illustrated this argument using several UV complete models as examples. While the MSSM, LQs, and also Little-Higgs-inspired models have problems with explaining both $a_e$ and $a_\mu$ simultaneously, models with vector-like fermions can account for both anomalies. Here, an Abelian flavor symmetry, for instance $L_\mu-L_\tau$, can be used to ensure the decoupling of the electron and the muon sector and even allow for intriguing connections to the anomalies observed in $b\to s\mu^+\mu^-$ transitions.

We stressed that considering heavy BSM degrees of freedom realized above the EW breaking scale requires chiral enhancement. The contributions to dipole moments in such models necessarily have a free phase, contrary to the non-enhanced effects which are real at the one-loop level. Therefore, in such a scenario, there is no reason to believe that the strict limit on the EDM of the electron should be reflected in the muon sector and a natural phase around $45^{\circ}$ would lead one to expect an EDM of $|d_\mu|\sim 3\times 10^{-22}\ecm$ at the current level of the $(g-2)_\mu$ anomaly. 
These models would remain viable if the tension in $a_e$ disappeared, but in this case explanations of $a_\mu$ that automatically lead to a small muon EDM would again become possible. 

While beyond the reach of the future $(g-2)_\mu$ experiments, an EDM at the level of $10^{-22}\ecm$ could be measured at the proposed muon EDM experiment at PSI using the frozen-spin technique,
with sensitivity to the phase of the relevant Wilson coefficient illustrated in Fig.~\ref{fig:sensitivity}. In combination with improved measurements of $a_e$, $a_\mu$, and the fine-structure constant $\alpha$, such an experiment would thus complete the search for BSM physics in lepton magnetic moments and, if the current anomalies were confirmed, provide further crucial insights into its flavor structure.

\begin{acknowledgments}
We thank Paul T.\ Debevec, David W.\ Hertzog, Klaus Kirch, and Adrian Signer for useful discussions. 
Financial support by the DOE (Grant No.\ DE-FG02-00ER41132) is gratefully acknowledged. A.C.\ is supported by an
Ambizione Grant of the Swiss National Science Foundation (PZ00P2\_154834).
\end{acknowledgments}

\end{document}